\documentclass[single-column]{ifacconf}

\usepackage{graphicx}      
\usepackage{natbib}        
\usepackage[dvipsnames]{xcolor}
\usepackage{amsmath}
\usepackage{amssymb}
\usepackage{mathrsfs} 
\usepackage{enumerate}
\usepackage{breqn}
\usepackage{tikz}
\usepackage{flushend}

\newcommand{\R}{\mathbb{R}}
\newcommand{\C}{\mathbb{C}}
\newcommand{\T}{\mathbb{T}}
\newcommand{\W}{\mathbb{W}}
\newcommand{\V}{\mathbb{V}}
\newcommand{\Lb}{\mathfrak{L}}
\newcommand{\B}{\mathfrak{B}}
\newcommand{\Bor}{\mathscr{B}}
\newcommand{\E}{\mathscr{E}}
\newcommand{\Z}{\mathbb{Z}}
\newcommand{\U}{\mathbb{U}}

\newcommand{\rk}{\mathrm{rk}}

\renewcommand{\l}{\ell}

\renewcommand{\ker}{\mathrm{Ker}}

\newcommand{\gia}{\color{black}}
\newcommand{\giar}{\color{black}}
\newcommand{\giarr}{\color{black}}

\makeatletter
\newcommand{\specialcell}[1]{\ifmeasuring@#1\else\omit$\displaystyle#1$\ignorespaces\fi}
\makeatother

\allowbreak

\begin{document}
\begin{frontmatter}

\title{\giarr LTI Stochastic Processes: \\ \ \  a Behavioral Perspective\thanksref{footnoteinfo}} 

\thanks[footnoteinfo]{The research leading to these results has received funding from the European Research Council under the Advanced ERC Grant Agreement Switchlet n. 670645.}

\author[First]{Giacomo Baggio} 
\author[Second]{Rodolphe Sepulchre} 

\address[First]{Dipartimento di Ingegneria dell'Informazione, Universit\`a di Padova,\\ via Gradenigo 6/B 35131 Padova, Italy.\\
       (e-mail: \texttt{giacomo.baggio@studenti.unipd.it}).}
\address[Second]{Department of Engineering, University of Cambridge\\ Trumpington Street, Cambridge CB2 1PZ, UK. \\ (e-mail: \texttt{r.sepulchre@eng.cam.ac.uk}).}

\begin{abstract}                
{\gia This paper revisits} the definition of linear time-invariant (LTI) stochastic process within a behavioral systems framework. Building on \cite{W:13}, we derive a canonical representation of an LTI stochastic process and a physically grounded notion of interconnection between independent stochastic processes. {\gia We use this framework to} analyze the {\giarr invariance properties enjoyed by distances between spectral densities of LTI processes.}
\end{abstract}


\end{frontmatter}

\section{Introduction}

{\gia Stochastic models play a crucial role in many areas of the natural and engineering sciences.} Indeed, mathematical models of stochastic phenomena are widely used in many branches of physics \citep{VK:07}, engineering \citep{A:70,LP:15}, economics \citep{Dupacova:02}, and biology \citep{Wilkinson:11}. {\giarr They aim at accounting for the fact that physical phenomena are unavoidably corrupted by some exogenous source of noise. This source of stochasticity sometimes represents an essential part of their model.} 
In this paper, we address the problem of modelling stochastic phenomena from an open systems viewpoint. {\gia A system is \emph{open} if it can interact with its external environment.} As opposed to closed systems, open systems are amenable to \emph{interconnection}. The term interconnection here is intended in the most general sense, i.e. as \emph{variable sharing} between systems \citep{W:07}. In this setting, a theory of open systems is de facto a theory of interconnected systems. When dealing with deterministic linear time-invariant (LTI, for short) systems, an elegant and mature theory of open systems is provided by the theory of behaviors \citep{WP:97}. {\gia The ultimate paper of Willems is a first step towards generalizing the latter theory to a stochastic framework. \cite{W:13} focuses on {\em static} stochastic systems.}


Building on \cite{W:13}, in this paper, we {\gia treat} the \emph{dynamical} case. We then specialize it to the case of discrete-time {\giarr LTI stochastic dynamical systems or, equivalently, LTI \emph{stochastic processes}}. In particular, we analyze: (i) a representation for such processes in terms of a linear model, (ii) in what sense and under which conditions these processes can be interconnected, and (iii) the {\giarr invariance properties naturally inherited by distances between these processes. Interestingly, once translated into the frequency domain, these invariance properties suggest to look for a natural class of projective-like distances between rational discrete-time spectral densities.}

{\em Paper structure.} The rest of the paper is organized as follows. After reviewing in \S2 some results of behavioral theory, in \S3 we introduce the {\gia behavioral definition of stochastic process}. In \S4 we focus on the linear-time invariant case. {\gia \S5 considers interconnection of LTI processes.} In \S6 we study the invariance properties {\giarr of distances between spectral densities of} LTI processes. Finally, \S7 collects some concluding remarks and future research directions.

{\em Notation. } Throughout the paper, we denote by $\Z$, $\R$, $\R_{+}$, and $\C$ the set of integer, real, non-negative real, and complex numbers, respectively. Given an element $A$ belonging to a set $\mathbb{A}$, $A^{\mathrm{c}}$ will denote the complement of $A$ w.r.t. $\mathbb{A}$, while $\emptyset$ the empty set. We let $\R^{n\times m}$ denote the set of real-valued $n\times m$ matrices and $A^{\top}\in\R^{m\times n}$ denote the transpose of $A\in\R^{n\times m}$. The symbols $\R[z,z^{-1}]^{n\times m}$ and $\R(z)^{n\times m}$ stand for the set of $n\times m$ Laurent polynomial matrices and the set of $n\times m$ rational matrices with real coefficients in the indeterminate $z$, respectively. The normal rank of a rational matrix $A(z)\in \R(z)^{n\times m}$ is defined as the rank of $A(z)$ almost everywhere in $z\in\C$ and it will be denoted by $\rk(A)$. Given $A(z)\in\R(z)^{n\times m}$, we let $A^{*}(z):=A^{\top}(1/z)$ and, for square $A(z)$'s of full normal rank, $A^{-*}(z):=[A^{\top}(1/z)]^{-1}$. We recall that a Laurent unimodular matrix is a square Laurent polynomial matrix whose inverse is also Laurent polynomial or, equivalently, whose determinant is a non-zero monomial $\lambda z^k$, $\lambda\in\R\setminus\{0\}$, $k\in\Z$. We denote by $\U[z,z^{-1}]^{n\times n}$ the group of $n\times n$ (Laurent) unimodular matrices and by $\mathcal{S}_+(\mathbf{T})^{n\times n}$ the set of $n\times n$ matrix-valued functions which are positive definite on the unit circle $\mathbf{T}:=\{\, z\in\C\,:\, |z|\ =1\, \}$, i.e. $n\times n$ discrete-time coercive spectral densities. 
$\mathcal{S}_{\mathrm{rat}}^{n\times n}\subset \mathcal{S}_+(\mathbf{T})^{n\times n}$ will denote the set of $n\times n$ \emph{rational} discrete-time coercive spectral densities. Finally, we shall suppose the reader to be acquainted with some elementary notions of probability theory, e.g. the definitions of $\sigma$-algebra, (smallest) $\sigma$-algebra generated by a collection of sets, Borel $\sigma$-algebra, probability measure; notions that can be found in any standard textbook of probability or measure theory, e.g.  \cite{B:86}.

\section{background on behavioral theory} \label{sec:notation}

In this preliminary section, we quickly review some basic notions and results of behavioral theory. We refer the reader to the seminal papers \cite{W:86,W:89,W:91} and to the monograph \cite{WP:97} for a comprehensive treatment on the subject.

In the theory of behaviors, a dynamical system is defined as a triple $\Sigma := (\T, \W, \B)$, where $\T$ is the set of times over which the system evolves (\emph{time axis}), $\W$ is the set over which the variables of the signals being modelled take values (\emph{signal space}), and $\B$ is a subset of $\W^{\T}$ (i.e. the set of all maps from $\T$ to $\W$, also called \emph{universum}) in which all the admissible system trajectories live (the \emph{behavior} of the system). 
The dynamical system  $\Sigma = (\T, \W, \B)$ is \emph{linear} if $\W$ is a vector space and $\B$ a linear subspace of $\W^{\T}$.
$\Sigma$ is said to be \emph{time-invariant} if $\T$ is closed under addition and $\sigma^t\B\subseteq \B$ for all $t\in\T$, where $\sigma$ denotes the backward shift operator defined as $(\sigma f)(t'):=f(t'+1)$.
In this paper we mainly focus on $n$-dimensional, real-valued, \emph{discrete-time} systems. Hence we set $\T:=\Z$ and $\W:=\R^{n}$. As a consequence, the behavior of the system becomes a subset of $(\R^{n})^{\Z}$ (the set of maps from $\Z$ to $\R^{n}$), i.e. a family of $n$-dimensional, real-valued, discrete-time sequences {\giar  $w\in(\R^{n})^{\Z}$.
An $n$-dimensional LTI system may be described by an Auto-Regressive (AR) model\footnote{We remark that the behavior is always \emph{deterministic}, i.e. it is composed by deterministic trajectories. Therefore, in this section, the term AR {\gia does not refer to} stochastic framework.} 
\[
	R_{\l}w(t+\l)+R_{\l+1}w(t+\l+1)+\dots+R_{L}w(t+L)=0,
\]
for all $t\in\Z$, where $R_{i}\in\R^{p\times n}$, $i=\l,\l+1,\dots,L$, $\l,L \in\Z,\, L>\l$}.
The Laurent polynomial matrix $R(z):=R_{\l}z^{\l}+R_{\l+1}z^{\l+1}+\dots+R_{L}z^{L}\in\R[z,z^{-1}]^{p\times n}$ defines an operator in the shift $\sigma$, $R\colon (\R^n)^\Z\to (\R^p)^\Z$ which allows to rewrite the previous expression as
\begin{align}\label{eq:ker-rep}
	R(\sigma)w(t)=0.
\end{align}
The behavior of the LTI system is then given by
\begin{align}\label{eq:ker-infty}
  \ker_{\infty} R :=\{w\in (\R^n)^\Z \, :\, R(\sigma)w=0\}.
\end{align}
Equation \eqref{eq:ker-rep} is known as the \emph{kernel representation} of an LTI behavior. 
An LTI behavior $\B\subseteq(\R^{n})^{\Z}$ is said to be \emph{complete} if for all $w(t)\in \B$, $\left.w\right|_{[t_{1},t_{2}]}\in \left.\B\right|_{[t_{1},t_{2}]}$, $\forall\, t_{1},t_{2}\in\Z$, $t_{1}\leq t_{2}$, where $\left.w\right|_{[t_{1},t_{2}]}$ and $\left.\B\right|_{[t_{1},t_{2}]}$ denote the restriction of $w(t)$ and $\B$, respectively, to the time interval $[t_{1},t_{2}]$. 
A fundamental result in behavioral theory states that every LTI complete behavior admits a kernel representation \eqref{eq:ker-rep}.

To conclude, consider two LTI complete behaviors $\B_{1}$, $\B_{2}$ with kernel representations described by $R_{1}(z)$, $R_{2}(z)\in\R[z,z^{-1}]^{p\times n}$, respectively. We say that the two behaviors $\B_{1}$, $\B_{2}$ are \emph{equivalent} if $R_{1}(z) = U (z) R_{2}(z)$ with $U(z)\in\U[z,z^{-1}]^{p\times p}$ being a Laurent unimodular matrix.  Hence, with reference to the kernel representation, every behavior is uniquely determined by its kernel matrix up to a unimodular transformation acting on the left. 

\section{From deterministic to \hspace{3cm} stochastic behaviors}

{\gia Using the terminology of \cite{W:13}, a \emph{stochastic system}
is defined} as a triple $(\V,\E,P)$, where $\V$ is the outcome space, $\E$ is a $\sigma$-algebra of events, and $P\colon \E\to [0,1]$ is a probability measure which assigns to each event in $\E$ a value in the interval $[0,1]$. 

Consider {\gia the} deterministic system $\Sigma=(\T,\W,\B)$. $\Sigma$ can be regarded as a very special {\gia stochastic system}. As a matter of fact, $\Sigma$ coincides with the {\gia stochastic system} $(\W^\T,\{\emptyset,\W^\T,\B,\B^{\mathrm{c}}\},P_\Sigma)$ {\giar in which $\B$ is a certain event, meaning that $P_{\Sigma}(\B)=1$.} Since the previous definition unavoidably involves the probability measure $P_\Sigma$ (and specifically the constraint $P_\Sigma(\B)=1$), we could clearly have used a $\sigma$-algebra of events richer than $\{\emptyset,\W^\T,\B,\B^{\mathrm{c}}\}$. However the latter seems to be a more natural choice due to the fact that it is the most ``parsimonious'' $\sigma$-algebra, that is, the smallest possible $\sigma$-algebra of events containing the deterministic behavior $\B$. 

Now assume that some source of stochasticity is added to the deterministic system $\Sigma$ (for instance, some additive noise acting on the trajectories of $\B$), then we expect that the newly generated (stochastic) system will possess a richer $\sigma$-algebra of events---indeed the noise modifies, and, more precisely, enlarges the space of admissible trajectories of the system---and, as a consequence, a new probability measure. Furthermore, by adding more and more sets to our event space, we are, in a sense, moving more and more away from the class of deterministic systems. Loosely speaking, the cardinality of the $\sigma$-algebra of events can be considered as a measure of the ``degree'' of stochasticity of the system. 

From this intuitive description, we can see that the $\sigma$-algebra of events plays an important role in the mathematical model of a stochastic system, perhaps as important as the probability measure associated to the system. When dealing with static systems, i.e. systems which do not evolve in time, this is exactly the point raised in \cite{W:13}. One of the aims of the present paper is to {\gia extend this viewpoint to} the dynamical case. To this end, we first revisit the definition of stochastic process in the spirit of behavioral theory.  

\begin{defn} A stochastic process is a quadruple $\Sigma:=(\T,\W,\E,P)$, where
\begin{enumerate}
\item $\T$ is the time axis,
\item $\W$ is the signal space, i.e. the set in which the variables whose (noisy) time evolution is modelled take on their values,
\item $\E$ is a $\sigma$-algebra of subsets of $\W^{\T}$ with elements called events,
\item $P\colon \E\to[0,1]$ is the probability measure defined on the $\sigma$-algebra of events.
\end{enumerate}
\end{defn}

With reference to the above definition, we observe that:
 \begin{enumerate}[(i)]
\item A stochastic process is a probability space where the outcome space is given by $\W^\T$. Two important classes of stochastic processes are obtained by selecting $\T=\R$, in which case the outcome space is the space of functions $f\colon \R\to \W$, and $\T=\Z$, in which case the outcome space is the space of sequences $\{f_t\}_{t\in\Z}$ taking values on $\W$. Intuitively, we can think of a stochastic process as a system described by a collection of ``behaviors'' {\giar with assigned probabilities, where $P$ specifies the probability of each ``behavior''.}
  
\item The standard definition of stochastic process is a family of random variables (i.e. measurable functions) $\{f_t\}_{t\in\T}$ defined on some probability space and parametrized by an index $t\in\T$, which usually represents time. By specifying the finite-dimensional probability distributions of the family $\{f_t\}_{t\in\T}$ it is then possible to characterize the infinite-dimensional distributions of the process (by virtue of Kolmogorov existence theorem \cite[Thm. 36.1]{B:86}). Our definition of a stochastic process is essentially equivalent to the latter one but formulated in terms of $\sigma$-algebras of events defined on the (usually infinite-dimensional) space of trajectories $\W^\T$. 
From this point of view, in Definition 1 emphasis is put on the event space itself rather than on the variables that generate that space.
\end{enumerate}

For the rest of the paper, we restrict our attention to the class of $n$-dimensional, real-valued, discrete-time stochastic processes, i.e. we set $\T=\Z$ and $\W=\R^n$. Notice that, in this setting, we can identify two particular subclasses of stochastic processes, namely the subclass of \emph{deterministic} dynamical systems whose $\sigma$-algebra of events is given by $\{\emptyset,(\R^n)^\Z,\B,\B^{\mathrm{c}}\}$ with $\B\subset(\R^n)^\Z$, and the subclass of \emph{classical} stochastic processes whose $\sigma$-algebra of events is given by the Borel $\sigma$-algebra generated by the open sets of $(\R^n)^\Z$ equipped with the product topology (i.e. the topology of pointwise convergence), which we will denote by $\Bor((\R^n)^\Z)$. Since the Borel $\sigma$-algebra of $(\R^n)^\Z$ coincides with the $\sigma$-algebra containing all the non-pathological subsets of $(\R^n)^\Z$, we can think of these two subclasses as two extremes in the space of stochastic processes.

\begin{rem}
  It is worth noting that when dealing with continuous-time stochastic processes ($\T=\R$) the Borel $\sigma$-algebra generated by the open sets of the product space $(\R^n)^\R$ equipped with the product topology proves often to be inadequate for describing the events of the process \cite[Ch.~7]{B:86}. Indeed, for instance, it can be shown that the many ``interesting'' sets of functions, e.g. the set of continuous functions, are not contained in $\Bor((\R^n)^\R)$. To overcome this issue, other types of $\sigma$-algebras can be considered in place of $\Bor((\R^n)^\R)$, for instance the Borel $\sigma$-algebra generated by the open sets in the space of continuous functions equipped the topology of uniform convergence on compact sub-intervals.\hfill$\diamondsuit$
\end{rem}

\section{Linear Time-Invariant \hspace{2cm}stochastic processes}

In this section we introduce the notion of linear time-invariant (discrete-time) stochastic process. {\giarr The events of these processes are uniquely defined up to trajectories belonging to a deterministic LTI behavior.}
We then discuss a canonical representation for these systems.

\begin{defn}\label{def:LTIproc}
The stochastic process $\Sigma:=(\Z,\R^{n},\E,P)$ is said to be \emph{linear} and \emph{time-invariant} (LTI, for short) if there exists a linear and time-invariant behavior $\Lb\subset(\R^{n})^{\Z}$ such that the events are the Borel subsets of the quotient space $(\R^{n})^{\Z}/ \Lb$, i.e. $\E:=\Bor((\R^{n})^{\Z}/ \Lb)$, and $P$ is a Borel probability measure on the same quotient space, i.e. $P\colon \Bor((\R^{n})^{\Z}/ \Lb) \to [0,1]$.
\end{defn}

Observe that $\Bor((\R^{n})^{\Z}/ \Lb)$ is a well-defined Borel $\sigma$-algebra. Indeed, it coincides exactly with the Borel $\sigma$-algebra generated by the open sets $A/\Lb$ (open sets of the quotient topology), with $A$ an open set of the topological vector space $(\R^n)^\Z$ equipped with the product topology. Moreover, using the terminology introduced in \cite{W:13}, we call $\Lb$ the \emph{fiber} of the LTI stochastic process. 

\begin{figure}[h!]
\begin{center}
\begin{tikzpicture}[
				scale=0.9, transform shape,
				>=latex,outer sep=0in,
				sample1/.style={draw=none,shape=rectangle,fill=gray,opacity=1,minimum width=1.2mm,minimum height=5mm,inner sep=0pt,outer sep=0pt}, 
				sample2/.style={draw=none,shape=rectangle,fill=gray,opacity=1,minimum width=1.2mm,minimum height=7.5mm,inner sep=0pt,outer sep=0pt},
				sample3/.style={draw=none,shape=rectangle,fill=gray,opacity=1,minimum width=1.2mm,minimum height=2.5mm,inner sep=0pt,outer sep=0pt}]

\draw[thick,->] (-0.5,0,0) -- (6,0,0) node[anchor=west]{$\Z$};
\draw[thick,->] (0,-0.5,0) -- (0,3.75,0) node[anchor=south]{$\R$};

\node[sample1] at (0,0.5) {};
\node[sample2] at (1,0.75) {};
\node[sample3] at (2,0.25) {};
\node[sample2] at (3,1.25) {};
\node[sample1] at (4,2) {};
\node[sample2] at (5,1.5) {};

\foreach \i in {1}
{
\node[sample1,red!50] at (0,0.5+\i) {};
\node[sample2,red!50] at (1,0.75+\i) {};
\node[sample3,red!50] at (2,0.25+\i) {};
\node[sample2,red!50] at (3,1.25+\i) {};
\node[sample1,red!50] at (4,2+\i) {};
\node[sample2,red!50] at (5,1.5+\i) {};
}



\draw[densely dashed,red] (-0.75,1,0) node[anchor=east]{$\ell\in\Lb$} -- (6,1,0);
\node[circle,fill=red,inner sep=0cm,minimum size=0.1cm] at (0,1) {};
\node[circle,fill=red,inner sep=0cm,minimum size=0.1cm] at (1,1) {};
\node[circle,fill=red,inner sep=0cm,minimum size=0.1cm] at (2,1) {};
\node[circle,fill=red,inner sep=0cm,minimum size=0.1cm] at (3,1) {};
\node[circle,fill=red,inner sep=0cm,minimum size=0.1cm] at (4,1) {};
\node[circle,fill=red,inner sep=0cm,minimum size=0.1cm] at (5,1) {};

\node[red!75] at (6,2.5) (E) {\small$E_{\ell}=\bar E+\ell$};


\node[fill=white,inner sep=0.05cm] at (5.4,1.5) {\small\textcolor{darkgray}{$\bar E$}};

\draw[|<->|,opacity=0.75] (4.25,2.25) -- (4.25,3.25);
\node[opacity=0.75] at (4.4,2.75){\footnotesize$\ell$};

\draw[dotted,opacity=1] (1,-0.2,0) node[anchor=north]{\scriptsize$1$} -- (1,3.4,0);
\draw[dotted,opacity=1] (2,-0.2,0) node[anchor=north]{\scriptsize$2$} -- (2,3.4,0);
\draw[dotted,opacity=1] (3,-0.2,0) node[anchor=north]{\scriptsize$3$} -- (3,3.4,0);
\draw[dotted,opacity=1] (4,-0.2,0) node[anchor=north]{\scriptsize$4$} -- (4,3.4,0);
\draw[dotted,opacity=1] (5,-0.2,0) node[anchor=north]{\scriptsize$5$} -- (5,3.4,0);

\end{tikzpicture}
\caption{\gia In an LTI stochastic process an event $E\in\E$ corresponds to a fixed event $\bar E$ plus all the subsets $E_{\ell}$ ``shifted'' by elements $\ell\in\Lb$. In this figure $\Lb$ is made of constant trajectories, i.e. trajectories belonging to the set $\{w\in\R^{\Z}\, :\, w(t+1)=w(t), \ \forall t\in\Z\}$.}
\label{Fig:2}
\end{center}
\end{figure}
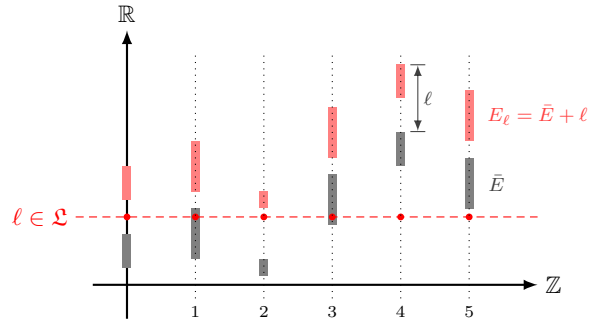

Definition \ref{def:LTIproc} can be intuitively interpreted as follows: given any (Borel) subset $\bar{E}\subset (\R^{n})^{\Z}$, which consists of a subset of trajectories in $(\R^{n})^{\Z}$, if the stochastic process $\Sigma$ is LTI with fiber $\Lb$ then the subset $E$ generated by adding to $\bar{E}$ the trajectories belonging to the LTI behavior $\Lb$ is an event of $\Sigma$. Loosely speaking, an event is a collection of subsets in $(\R^{n})^{\Z}$, {\giar where each subset contains} trajectories ``parallel'' to the LTI behavior $\Lb$ (see Fig.~\ref{Fig:2} for {\gia an illustration}).

Throughout the paper, we will often make the assumption that fibers are \emph{complete} LTI behaviors, this means that all the trajectories belonging to the fiber admit a finite dimensional characterization, that is, they are uniquely determined by their restrictions over all possible finite time intervals. Under this assumption, the fiber of an LTI stochastic process can always be represented by means of the kernel of a Laurent polynomial matrix, as recalled in \S\ref{sec:notation}. LTI processes characterized by a complete fiber admit a canonical representation, called \emph{kernel representation} by analogy with the deterministic case.

\begin{thm}\label{thm:ker-rep}
 A stochastic process $\Sigma =(\Z,\R^n,\E,P)$ is an LTI stochastic process with fiber $\Lb$ being a complete LTI behavior if and only if $\Sigma$ can be described by a stochastic sequence $w(\cdot)$ satisfying for all $t\in\Z$
 \begin{equation}\label{eq:ARdyn}
	R(\sigma)w(t)=e(t), 
 \end{equation}
where $R(z)\in\R[z,z^{-1}]^{m\times n}$ is of full row normal rank, i.e. $\rk(R)=m$, and $e(\cdot)$ describes the stochastic process $\Sigma_e:=(\Z,\R^m,\Bor((\R^m)^\Z), P_e)$.
\end{thm}

\begin{pf}
  See Appendix \ref{sec:app}.
\end{pf}

With reference to the previous result, we remark that if $E\in \E$ is an event of $\Sigma$ then its probability measure $P$ is defined through $e(\cdot)$ by
\[
  P(E):=P_e(R[E]),
\]
being $R[E]\in\Bor((\R^m)^\Z)$ the image of $E$ under $R(\sigma)$.


Furthermore, if we consider the restriction of an LTI process to a finite set of time indices, say $I:=\{t_1,t_2,\dots,t_n\}$, $t_i\in\Z$, $n> 0$, then we obtain a (static) stochastic system described by the triple $\Sigma|_I := ((\R^n)^{|I|},\E|_I,P|_I)$, where $|I|$ is the cardinality of the set $I$, $\E|_I:=\Bor((\R^n)^{|I|}/\Lb|_I)$ with $\Lb|_I$ the restriction of the complete LTI behavior to the time set $I$, and $P|_I$ a restricted probability measure defined for all $E\in\E|_I$ as 
\[
	\textstyle P|_I(E):=P\left(\bigcup_{G_i\in\pi^{-1}[E]}G_i\right),
\]
 being $\pi^{-1}[E]$ the pre-image of $E$ under the canonical projection $\pi\colon (\R^n)^\Z\to (\R^n)^{|I|}$, $\{f_t\}_{t\in\Z} \mapsto \{f_{t_1},f_{t_2},\dots,f_{t_n}\}$. Since, in general, the restriction $\Lb|_I$ returns a non-empty linear finite-dimensional subspace of $(\R^n)^{|I|}$, we note that $\Sigma|_I$ does not, in general, describe a \emph{classical} random vector, where for classical we mean a random vector characterized by a Borel $\sigma$-algebra of events on $(\R^n)^{|I|}$, as in \cite[Def. 2]{W:13}.

\section{interconnection of \hspace{3cm} stochastic processes}
Interconnection is a property characterizing open systems, i.e. systems which are allowed to interact with their environment. With reference to the mathematical model of a deterministic dynamical system, this interaction can take place if some variables of the system are left unmodelled \citep{W:07}. 
In this section we present an extension of the definition of interconnection between deterministic dynamical systems which applies to stochastic processes. After introducing some general definitions, we will focus on the discrete-time LTI case.

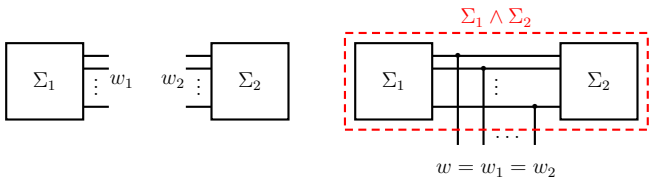
\begin{figure}[h!]
\begin{center}
\begin{tikzpicture}[scale=0.675, transform shape,>=latex,outer sep=0in]

\begin{scope}[xshift=-1.8cm]
\draw[thick] (0,0) rectangle (1.5,1.5);
\draw[thick] (4,0) rectangle (5.5,1.5);
\node at (0.75,0.75) {\large $\Sigma_{1}$};
\node at (4.75,0.75) {\large $\Sigma_{2}$};
\draw[thick] (1.5,1.25) -- (2,1.25);
\draw[thick] (1.5,1) -- (2,1);
\draw[thick] (1.5,0.25) -- (2,0.25);
\node at (1.75,0.75) {\large $\vdots$};
\node at (2.25,0.75) {\large $w_{1}$};

\draw[thick] (3.5,1.25) -- (4,1.25);
\draw[thick] (3.5,1) -- (4,1);
\draw[thick] (3.5,0.25) -- (4,0.25);
\node at (3.75,0.75) {\large $\vdots$};
\node at (3.25,0.75) {\large $w_{2}$};
\end{scope}

\begin{scope}[xshift=5cm,yshift=2.5cm]
\draw[thick] (0,-2.5) rectangle (1.5,-1);
\draw[thick] (4,-2.5) rectangle (5.5,-1);
\node at (0.75,-1.75) {\large $\Sigma_{1}$};
\node at (4.75,-1.75) {\large $\Sigma_{2}$};
\draw[thick] (1.5,-1.25) -- (4,-1.25);
\draw[thick] (1.5,-1.5) -- (4,-1.5);
\draw[thick] (1.5,-2.25) -- (4,-2.25);
\node at (2.75,-1.75) {\large $\vdots$};

\draw[thick] (2,-1.25) node {\tiny$\bullet$} -- (2,-3);
\draw[thick] (2.5,-1.5) node {\tiny$\bullet$} -- (2.5,-3);
\node at (3,-2.85) {\large  $\dots$};
\draw[thick] (3.5,-2.25) node {\tiny$\bullet$} -- (3.5,-3);
\node at (2.75,-3.5) {\large $w=w_1=w_2$};

\draw[red,thick,densely dashed] (-0.2,-2.7) rectangle (5.7,-0.8);
\node at (2.75,-0.5) {\large \color{red}$\Sigma_{1}\wedge \Sigma_{2}$};
\end{scope}
\end{tikzpicture}
\caption{Interconnection of stochastic processes $\Sigma_1=(\T,\W,\E_1,P_1)$ and $\Sigma_2=(\T,\W,\E_2,P_2)$ .}
\label{Fig:interc}
\end{center}
\end{figure}

As in the deterministic case, interconnection of two stochastic processes can be thought of as \emph{variable sharing} between the two processes. In other words, interconnection between two processes is obtained by simply imposing an equality constraint on the variables describing the stochastic laws of the two processes (Fig.~\ref{Fig:interc}).

In the deterministic case, given two dynamical systems $\Sigma_1=(\W,\T,\B_1)$ and $\Sigma_2=(\W,\T,\B_2)$ having the same time axis and signal space, the interconnection between $\Sigma_1$ and $\Sigma_2$ is defined as the deterministic system $\Sigma_1\wedge \Sigma_2:=(\W,\T,\B_1\cap \B_2)$ \citep{W:07}. In the stochastic case, the definition of interconnection we are going to present is similar to the latter one if we replace the role of the deterministic behaviors with the $\sigma$-algebras of events of the processes (which indeed represent collections of admissible ``behaviors'' of the processes). However, in this case, a problem arises. As a matter of fact, since interconnection of stochastic processes also involves the probability laws defined on the processes, a natural compatibility condition between the two to-be-interconnected processes has to be fulfilled. This natural condition states that the probability measure defined on the interconnected process must be consistent, in a sense explained below, with the probability measures defined on the original processes. This condition was introduced in \cite{W:13} with reference to (static) stochastic systems under the name of \emph{complementarity}. In the following definition, we adapt the notion of complementarity to the case of stochastic processes.

\begin{defn}
  Two stochastic processes $\Sigma_{1}=(\T,\W,\E_{1},P_{1})$ and $\Sigma_{2}=(\T,\W,\E_{2},P_{2})$ are said to be \emph{complementary} if for all $E_{1}$, $E_{1}'\in\E_{1}$ and $E_{2}$, $E_{2}'\in\E_{2}$ such that $E_{1}\cap E_{2}=E_{1}'\cap E_{2}'$ it holds
\[
  P_{1}(E_{1})P_{2}(E_{2})=P_{1}(E_{1}')P_{2}(E_{2}').
\]
Moreover, the two $\sigma$-algebras $\E_1$ and $\E_2$ are said to be \emph{complementary} if for all non-empty sets $E_{1}$, $E_{1}'\in\E_{1}$ and $E_{2}$, $E_{2}'\in\E_{2}$ such that $E_{1}\cap E_{2}=E_{1}'\cap E_{2}'$ it holds
\[
  E_{1}=E_{1}'\text{ and } E_{2}=E_{2}'.
\]
\end{defn}

\begin{rem}
The notion of complementarity between $\sigma$-algebra of the processes is weaker than the notion of complementarity of processes. Indeed, the former represents only a sufficient condition for complementarity of processes. However, working with complementarity of $\sigma$-algebra is usually easier since this notion does not involve the probability laws describing the processes, as pointed out also in \cite{W:13}. \hfill$\diamondsuit$
\end{rem}

Under the assumption of complementarity between two stochastic processes, we arrive at a formal definition of interconnection.

\begin{defn}\label{def:interc}
Let $\Sigma_{1}=(\T,\W,\E_{1},P_{1})$ and $\Sigma_{2}=(\T,\W,\E_{2},$ $P_{2})$ be two independent\footnote{We say that two stochastic processes are (stochastically) independent if their $\sigma$-algebras of events are so, with respect to any joint probability measure.} and complementary stochastic processes. The \emph{interconnection} of $\Sigma_{1}$ and $\Sigma_{2}$ is defined as the stochastic process
\[
	\Sigma_{1}\wedge \Sigma_{2}:=(\T,\W,\E,P),
\]
where $\E$ is the $\sigma$-algebra generated by $\E_{1}\cup \E_{2}$ and $P$ is defined on the sets $\{E_{1}\cap E_{2}\,:\, E_{1}\in\E_{1}, E_{2}\in \E_{2}\}$ by 
\[
	P(E_{1}\cap E_{2}):= P_{1}(E_{1})P_{2}(E_{1})
\]
and extended to all of $\E$ by virtue of the Hahn-Kolmogorov extension theorem  \cite[Ch.~3]{B:86}.\footnote{The Hahn-Kolmogorov theorem gives conditions under which a function $\mu\colon \mathscr{A}\to [0,1]$ defined on an algebra $\mathscr{A}$ of subsets of $\Omega$ can be extended to a unique bona fide probability measure on the $\sigma$-algebra generated by $\mathscr{A}$. These conditions are: (i) $\mu(\Omega)=1$,  (ii) countably additivity, i.e. $\mu\left(\bigcup_{i=1}^\infty A_i\right)=\sum_{i=1}^\infty \mu(A_i)$ for any countable disjoint family of subsets $\{A_i\}_{i=1}^\infty$, $A_i\in\mathscr{A}$, such that $\bigcup_{i=1}^\infty A_i\in\mathscr{A}$. (A function satisfying these two requirements is called a \emph{pre-measure} on $\mathscr{A}$.) In our case, the theorem applies to $P$ since the latter is defined through the product of two bona fide probability measures.}
\end{defn}

\begin{rem}
  It is worth pointing out that interconnection between stochastic processes, as given in Definition \ref{def:interc}, differs from the classical notion of \emph{coupling} between stochastic processes. As a matter of fact, even in the static case, coupling of two stochastic systems $\Sigma_1=(\W,\E_{1},P_{1})$ and $\Sigma_2=(\W,\E_{2},P_{2})$ requires the construction of a new stochastic system with signal space $\W\times \W$, $\sigma$-algebra generated by the sets $\{E_1\times E_2\,:\, E_{1}\in\E_{1},\, E_{2}\in\E_{2}\}$, and probability measure having prescribed marginal distributions, see e.g. \cite{L:02}. On the other hand, interconnection between $\Sigma_1$ and $\Sigma_2$ means that a new $\sigma$-algebra is constructed on the \emph{same} signal space shared by the two to-be-interconnected systems, that is $\W$. More precisely, the events which lie in the intersection between the two $\sigma$-algebras $\E_{1}$ and $\E_{2}$ generate the $\sigma$-algebra of the interconnected system. From this viewpoint, coupling appears more similar to \emph{juxtaposition} of stochastic processes than interconnection, where for juxtaposition we mean that starting from two processes described by stochastic laws $w_1$ and $w_2$ we construct a new process described by $(w_1,w_2)$, as in \cite{W:13}. {\gia We can therefore think of interconnection as a \emph{restriction} in the signal space and coupling as an \emph{expansion} of the signal space.}\hfill$\diamondsuit$
\end{rem}

We now restrict the attention to the class of LTI stochastic processes. In this case, the fiber of the process is given by an LTI behavior. If we add the further assumption that the fiber is described by a complete LTI behavior, then, by virtue of Theorem \ref{thm:ker-rep}, the process admits a kernel representation. For this class of stochastic processes, it is possible to derive a condition on the kernel matrices which is equivalent to complementary of $\sigma$-algebras of events of the processes.

\begin{thm}\label{thm:interc}
  Consider two (stochastically) independent LTI complete stochastic processes $\Sigma_{1}:=(\Z,\R^{n},\E_{1},P_{1})$ and $\Sigma_{2}:=(\Z,\R^{n},\E_{2},P_{2})$ described by fibers $\Lb_1:=\ker_\infty R_1$ and $\Lb_2:=\ker_\infty R_2 $, for suitable Laurent polynomial matrices $R_{1}(z)\in\R[z,z^{-1}]^{m\times n}$, $R_{2}(z)\in\R[z,z^{-1}]^{p\times n}$ with $\rk(R_{1})=m$ and $\rk(R_{2})=p$. The two $\sigma$-algebras $\E_{1}$ and $\E_{2}$ are complementary if and only if it holds
  \begin{align}\label{eq:rank_condition}
    \mathrm{rk}\begin{bmatrix}R_1 \\ R_2 \end{bmatrix} = m+p.
  \end{align}
In this case, the fiber of the interconnected process $\Sigma_1\wedge\Sigma_2$ is given by 
\begin{align}
  \Lb_{1\wedge 2}:=\Lb_1\cap \Lb_2=\ker_\infty\begin{bmatrix}R_1 \\ R_2 \end{bmatrix}.
\end{align}
\end{thm}

\begin{pf}
  See Appendix \ref{sec:app}.
\end{pf}

\begin{exmp}\label{ex:interc}
As a simple example of interconnection, consider two LTI processes $\Sigma_1:=(\Z,\R^2,\E_1,P_1)$ and $\Sigma_2:=(\Z,\R^2,\E_2,P_2)$ described by kernel representations
\begin{align*}
  \begin{bmatrix}\sigma+a_1 & \sigma+b_1\end{bmatrix} w_1(t) &= e_1(t),\ \ \ a_1,\, b_1\in\R,\\
  \begin{bmatrix}\sigma+b_2 & \sigma+a_2\end{bmatrix} w_2(t) &= e_2(t),\ \ \ a_2,\, b_2\in\R,
\end{align*}
respectively. Furthermore, assume that $e_1(\cdot)$ and $e_2(\cdot)$ describe stochastically independent processes $\Sigma_{e_1}=(\Z,\R,\Bor(\R^\Z),P_{e_1})$ and $\Sigma_{e_2}=(\Z,\R,\Bor(\R^\Z),P_{e_2})$, respectively. If we partition the variables $w_1$ and $w_2$ in a ``input-output'' form $w_1:=[u_1\ y_1]^\top$ and $w_2:=[y_2\ u_2]^\top$, then the two LTI processes $\Sigma_1$ and $\Sigma_2$ can be regarded as two noisy input/output LTI systems (see also Fig.~\ref{Fig:interc_ex}). 
The two $\sigma$-algebras $\E_1$ and $\E_2$ are complementary if and only if
\[
  \mathrm{rk}\begin{bmatrix}z+a_1 & z+b_1 \\ z+b_2 & z+a_2 \end{bmatrix} = 2,
\]
or, equivalently, if and only if $a_1+a_2\neq b_1+b_2$ and $a_1a_2\neq b_1b_2$. If the latter conditions are met, the interconnected process $\Sigma_1\wedge \Sigma_2$ is a well-defined LTI process described by the laws of the stochastic sequence $w(t):=[w_{1}(t)\ w_{2}(t)]^{\top}$ satisfying {\giar
\[
  \begin{bmatrix}\sigma+a_1 & \sigma+b_1 \\ \sigma+b_2 & \sigma+a_2 \end{bmatrix} w(t) = e(t),
\]
where $e(\cdot)$ describe the process $\Sigma_{e}=(\Z,\R^{2},\Bor((\R^2)^\Z),P_{e})$ with $P_{e}$ defined as $P_{e}(E_{1}\times E_{2}):=P_{e_{1}}(E_{1})P_{e_{2}}(E_{2})$ for all $E_{1},\, E_{2}\in \Bor(\R^\Z)$ and extended to all of $\Bor((\R^2)^\Z)$ via the Hahn-Kolmogorov extension theorem.
}\hfill$\diamondsuit$

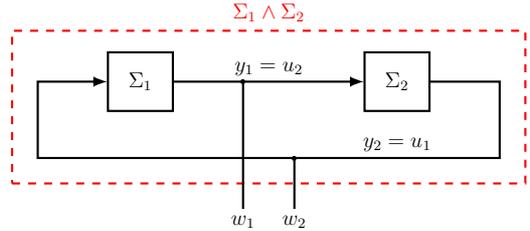
\begin{figure}
\begin{center}
\begin{tikzpicture}[scale=0.675, transform shape]

  \node[draw,thick,inner sep=0.4cm] (Tone) at (0,-2.5) {\large  $\Sigma_{1}$};
  \node[draw,thick,inner sep=0.4cm] (Ttwo) at (5,-2.5) {\large $\Sigma_{2}$};
  \draw[thick,-latex] (Tone) -- (Ttwo) node[above,midway]{\large $y_1=u_2$};
  \draw[thick,-latex] (Ttwo) -| (7,-4) -- (-2,-4) node[above,midway]{\large  \hspace{5cm}$y_2=u_1$} |- (Tone);;
  \node[red] at (2.5,-1.15) {\large $\Sigma_1\wedge\Sigma_2$};
  \draw[dashed,red,thick] (-2.5,-1.5) -- (7.5,-1.5) -- (7.5,-4.5) -- (-2.5,-4.5) -- cycle;

  \node (A) at (2,-2.5) {\tiny $\bullet$};
  \node (B) at (3,-4) {\tiny $\bullet$};
  \draw[thick] (2,-2.5) -- (2,-5) node[below] {\large $w_1$};
  \draw[thick] (3,-4) -- (3,-5) node[below] {\large $w_2$};;

\end{tikzpicture}
\caption{Interconnection of LTI stochastic processes $\Sigma_1$ and $\Sigma_2$ of Example \ref{ex:interc} in an input-output representation.}
\label{Fig:interc_ex}
\end{center}
\end{figure}
\end{exmp}

To conclude this section, we present a straightforward corollary of Theorem \ref{thm:interc} which gives a characterization of interconnected processes with a ``full'' $\sigma$-algebra of events.

\begin{cor}
 Consider the two Laurent polynomial matrices $R_1(z)\in\R[z,z^{-1}]^{m\times n}$ and $R_2(z)\in\R[z,z^{-1}]^{p\times n}$ describing the LTI processes defined in Theorem \ref{thm:interc}. The $\sigma$-algebra of the interconnected system $\Sigma_1\wedge \Sigma_2$ is given by $\E = \Bor((\R^n)^\Z)$ if and only if $R:=[R_1^{\top} \ R_2^{\top}]^\top\in\U[z,z^{-1}]^{n\times n}$, i.e. $R$ is a unimodular Laurent polynomial matrix.
\end{cor}

{\giarr
\section{Metric invariance properties of LTI stochastic  processes}
}


In the previous sections, we have pointed out that in the definition of stochastic process a crucial role is played by the event space $\E$. For LTI stochastic processes, the structure of the event space is characterized by its \emph{fiber}, i.e. by the subspace of trajectories described by an LTI behavior.  Let us restrict the attention to fibers defined by complete behaviors and consider two stochastic processes in their kernel representations
\begin{align}
	\Sigma_{1}\colon\quad R_1(\sigma)w_1(t) &= e_{1}(t), \label{eq:w1}\\
	\Sigma_{2}\colon\quad R_2(\sigma)w_2(t) &= e_{2}(t). \label{eq:w2}
\end{align}
with $R_{1}(z)\in\R[z,z^{-1}]^{m\times n}$, $R_{2}(z)\in\R[z,z^{-1}]^{m\times n}$, $\rk(R_{1})=\rk(R_{2})=m$, and $e_1$, $e_{2}$ describing the processes $(\Z,\R^m,\Bor((\R^m)^\Z),P_{e_1})$, $(\Z,\R^m,\Bor((\R^m)^\Z),P_{e_2})$, respectively.
If the two processes happen to have the same fiber, then it follows that $\ker_{\infty}R_{1}$ and $\ker_{\infty}R_{2}$ are equivalent behaviors. This in turn implies that $R_{1}$ and $R_{2}$ are connected by a unimodular transformation acting on the left, i.e. $R_{1}(z)=U(z)R_{2}(z)$ with $U(z)\in\U[z,z^{-1}]^{m\times m}$. {\giarr In view of this fact, we introduce the following definition of equivalence between complete LTI stochastic processes.

\begin{defn} Consider two complete LTI stochastic processes $\Sigma_{1}$, $\Sigma_{2}$ described by kernel matrices $R_{1}$, $R_{2}\in \R[z,z^{-1}]^{m\times n}$. We say that $\Sigma_{1}$ and $\Sigma_{2}$  are \emph{equivalent} if their fibers are equivalent behaviors, or, equivalently, if $R_{1}=U R_{2}$ with $U\in\U[z,z^{-1}]^{m\times m}$. 
\end{defn}
}

We now investigate what the {\giarr above defined equivalence between LTI processes entails, when considering \emph{spectral densities} of LTI stochastic processes.} To this extent, assume that $e_{1},\,e_{2}$ are unit-variance white noise processes, and $w_{1}$, $w_{2}$ stochastic processes in the classical sense, i.e. equipped with Borel $\sigma$-algebras of events.\footnote{{\gia The reader will notice that the classical notion of stochastic process is always recovered} by enriching the $\sigma$-algebra of events and suitably redefining the probability over the new $\sigma$-algebra.} For $n=m$, the spectral densities of $w_{1}$ and $w_{2}$ in \eqref{eq:w1}-\eqref{eq:w2} are given, respectively, by
\[
	\Phi_{1}(z):=R_{1}^{-1}(z)R_{1}^{-*}(z), \quad \Phi_{2}(z):=R_{2}^{-1}(z)R_{2}^{-*}(z),
\]
and the fact that $\Sigma_{1}$ and $\Sigma_{2}$ are equivalent translates into the relation
\[
	 R_{1}^{-1}(z)=R_{2}^{-1}(z)V(z),
\]
for a suitable unimodular matrix $V(z)\in\U[z,z^{-1}]^{n\times n}$. 
{{\giarr
We can generalize this observation to arbitrary rational discrete-time coercive spectral densities as follows.

\begin{defn}\label{def:equivalencesd}
Consider two rational spectral densities $\Phi_{1}, \, \Phi_{2}\in \mathcal{S}_{\mathrm{rat}}^{n\times n}$ and let $W_{1},\, W_{2} \in\R(z)^{n\times n}$ be spectral factors of $\Phi_{1}, \, \Phi_{2}$, respectively, that is $\Phi_{i}=W_{i}W_{i}^{*}$, $i=1,2$. We say that $\Phi_{1}, \, \Phi_{2}$ are \emph{unimodular equivalent} if 
\begin{align}\label{eq:uniminv-2}
	W_{1}(z) = W_{2}(z) V(z),
\end{align}
with $V(z)\in\U[z,z^{-1}]^{n\times n}$.
\end{defn}

\begin{rem}\label{rem:equivalencesd}
Notice that the above notion of unimodular equivalence depends upon the choice of the spectral factors $W_{1}$ and $W_{2}$ of $\Phi_{1}$ and $\Phi_{2}$, respectively. However, if we restrict $W_{1}$ and $W_{2}$ to belong to the set of stable spectral factors with stable inverse, i.e. spectral factors analytic in the region $\mathcal{D}:=\{\,z\in\C\,:\,|z|\ >1\,\}$ with inverse analytic in the same region, then this is not the case.\footnote{\giarr Here the term \emph{stable} refers to the \emph{finite} zero-pole structure of a spectral factor. In particular, such a spectral factor can possess zeros/poles at infinity.}
This follows from the fact that 
\begin{enumerate}[(i)]
\item if $W',\,W''\in \R(z)^{n\times n}$ are stable spectral factors of $\Phi\in \mathcal{S}_{\mathrm{rat}}^{n\times n}$ with stable inverse, then $W' = W'' Q$ with $Q\in\R(z)^{n\times n}$ being an all-pass (Laurent) unimodular matrix;
\item product of (Laurent) unimodular matrices is again (Laurent) unimodular.
\end{enumerate}
In view of this, in Definition \ref{def:equivalencesd} we will always consider spectral factors belonging to the set
\begin{flalign*}
	\quad\ \  \mathscr{W}:=\{\,W\in\R(z)^{n\times n}\,:\ \, & \Phi=WW^{*} \text{ and } \\
	& W,\ W^{-1} \text{ analytic in } \mathcal{D} \,\}. && \diamondsuit
\end{flalign*}
\end{rem}

In light of Definition \ref{def:equivalencesd} and Remark \ref{rem:equivalencesd}, a spectral density $\Phi\in\mathcal{S}_{\mathrm{rat}}^{n\times n}$ is uniquely determined by any spectral factor in $\mathscr{W}$, modulo a unimodular matrix.

At this point, it is interesting to analyze the invariance properties of distances in $\mathcal{S}_{\mathrm{rat}}^{n\times n}$, naturally induced by the unimodular equivalence relation \eqref{eq:uniminv-2}. Specifically, from \eqref{eq:uniminv-2} it follows that any natural distance $d\colon \mathcal{S}_{\mathrm{rat}}^{n\times n} \times \mathcal{S}_{\mathrm{rat}}^{n\times n}\to \R_{+}$ must satisfy
\begin{align}
d(\Phi_{1},\Phi_{2}) & = d(W_{1}W_{1}^{*},W_{2}W_{2}^{*}) \notag \\
& =  d(W_{1}V_{1}V_{1}^{*}W_{1}^{*},W_{2}V_{2}V_{2}^{*}W_{2}^{*}), \label{eq:invariance}
\end{align}
for unimodular matrices $V_{1}$, $V_{2}\in\U[z,z^{-1}]^{n\times n}$. We term the invariance property \eqref{eq:invariance} \emph{unimodular invariance}. The behavioral  framework suggests that distances between spectral densities of LTI stochastic processes should be unimodular invariant. 

In particular, w.r.t. the scalar case, unimodular transformations take the form $u(z)=\lambda z^{k}$, $\lambda\in\R\setminus\{0\}$ and $k\in\Z$. Therefore unimodular invariance \eqref{eq:invariance} reduces to a projective invariance property. Namely, letting $\Phi_{1}, \, \Phi_{2}\in \mathcal{S}_{\mathrm{rat}}^{1\times 1}$, it holds
\begin{align*}
	d(\Phi_{1},\Phi_{2})=d(\alpha_{1}\Phi_{1}, \alpha_{2}\Phi_{2}),\quad \alpha_{1},\, \alpha_{2}>0.
\end{align*}
Consequently, in the scalar case, a distance which satisfies unimodular invariance \eqref{eq:invariance} is a distance between ``shapes'' of spectral densities, since such a distance must be insensitive to scalings. Distances of this kind have been extensively investigated in the literature, see e.g. \cite{M:00,G:07b,G:07}, and have found interesting applications in the area of speech processing. 
{\giarr
With reference to the multivariate case, it would be interesting to investigate the existence of metrics that are invariant in the sense of Eq.~\eqref{eq:invariance}.
This topic will be the subject of future investigation.
}
}

\section{Conclusions and future work}

In this paper, we addressed the problem of modelling stochastic dynamical systems from a behavioral perspective. {\giarr We focused on LTI processes and we analyzed their interconnection and metric invariance properties. More specifically, an open-systems notion of interconnection between LTI stochastic processes and a natural invariance property enjoyed by distances in the space of rational spectral densities have been discussed. Directions for future research include: (i) the analysis of the notion of interconnection here introduced and its applications, (ii) the study of metrics in the space of rational discrete-time spectral densities featuring the invariance property presented in \S6.}

\bibliography{Bibliography}             


\appendix
\section{Extended proofs}\label{sec:app}	    	
                                                                    
In this Appendix, we present the proofs of Theorem \ref{thm:ker-rep} and Theorem \ref{thm:interc} of the main text.

\textbf{Proof of Theorem \ref{thm:ker-rep}.} \  \emph{``If''}: Assume that the stochastic process $\Sigma$ is described by the stochastic law of the sequence $w(\cdot)$ satisfying \eqref{eq:ARdyn}. We first recall some facts concerning the topological vector space $(\R^n)^\Z$ and polynomial operators in the shift, which can be found in \cite[\S4]{W:89}. The space of time series $(\R^n)^\Z$ when equipped with the product topology is a completely metrizable and separable (i.e. Polish) topological vector space. Also, $\Lb:=\ker_{\infty} R$ is a closed and linear subspace of $(\R^n)^\Z$. The polynomial operator in the shift $R(\sigma)$ is a linear, continuous, and surjective (since of full row normal rank) operator from $(\R^n)^\Z$ to $(\R^m)^\Z$.  Consider now the quotient space $(\R^n)^\Z/\Lb$. Since $\Lb$ is a closed and linear subspace of $(\R^n)^\Z$, this is again a Polish space, as it is separable (as every quotient space of a separable space) and completely metrizable w.r.t. the induced quotient topology (see e.g. \cite[Ch.1 \S3.2]{Bourbaki}). Hence by taking the restriction of $R(\sigma)$ to the quotient space $(\R^n)^\Z/\Lb$, i.e. $R|_{(\R^n)^\Z/\Lb}$, we obtain a continuous and bijective operator between Polish spaces. From this fact it follows from \cite[Thm. 15.1]{Kechris:12} that $R|_{(\R^n)^\Z/\Lb}$ is a Borel isomorphism, i.e. both $R|_{(\R^n)^\Z/\Lb}$ and its inverse are Borel measurable. This implies that to each event set $E_{e}\in\Bor((\R^m)^\Z)$ with associated probability $P_{e}(E_{e})$ corresponds one and only one $E:=R^{-1}[E_{e}]\in\Bor((\R^n)^\Z/\Lb)$ with probability $P(E):=P_{e}(E_{e})$, where $R^{-1}[A]$ denotes the pre-image of $A$ under $R(\sigma)$. Hence $w(\cdot)$ defines the LTI stochastic process $(\Z,\R^n,\Bor((\R^n)^\Z/\Lb),P)$, with $\Lb=\ker_{\infty} R$ being a complete LTI behavior.
  
  \emph{``Only if''}: Assume now that $\Lb$ is a complete LTI behavior and $\Sigma=(\Z,\R^{n},\E,P)$ an LTI process with fiber $\Lb$. Since $\Lb$ is complete, there exists a Laurent polynomial matrix $R(z)\in\R[z,z^{-1}]^{m\times n}$,  $\rk(R)=m\leq n$, such that $\Lb=\ker_\infty R$ \cite[\S4]{W:89}. As before, $R(\sigma)$ restricted to $(\R^n)^\Z/\Lb$ is a Borel isomorphism between $(\R^n)^\Z/\Lb$, equipped with the quotient topology, and $(\R^m)^\Z$, equipped with the product topology. Consider
  \[
    R(\sigma)w(t)=e(t),
  \]
  where $e(\cdot)$ describes a stochastic process with signal space $\R^m$, $\sigma$-algebra $\Bor((\R^m)^\Z)$, and probability $P_{e}$ such that $P_{e}(R[E]):=P(E)$ for all $E\in \Bor((\R^n)^\Z/\Lb)$, being $R[E]\in\Bor((\R^m)^\Z)$ the image of $E$ under $R$. From this construction it follows that the LTI stochastic process $((\R^n)^\Z,\Bor((\R^n)^\Z/\Lb),P)$ is described by the stochastic law of $w(\cdot)$ in \eqref{eq:ARdyn}. This concludes the proof. \qed  

\textbf{Proof of Theorem \ref{thm:interc}.} \ 
  \emph{``If''}: Assume that the normal rank condition in \eqref{eq:rank_condition} holds. Firstly, observe that this implies that $m+p\leq n$, otherwise equality in \eqref{eq:rank_condition} can not be attained. Secondly, as noticed in the proof of Theorem \ref{thm:ker-rep}, the operator $R_{1}$, when restricted to the domain $(\R^{n})^{\Z}/\ker_{\infty} R_{1}$, is a Borel isomorphism between topological spaces $(\R^{n})^{\Z}/\ker_{\infty} R_{1}$ and $(\R^{m})^{\Z}$. A similar result holds for $R_{2}$. This implies that $R_{1}$ ($R_{2}$, respectively) establishes a one-to-one correspondence between Borel sets in $(\R^{m})^{\Z}$ ($(\R^{p})^{\Z}$) and Borel sets in $(\R^{n})^{\Z}/\ker_{\infty} R_{1}$ ($(\R^{n})^{\Z}/\ker_{\infty} R_{2}$). Therefore every Borel set $\bar{E}_1\in\Bor((\R^{m})^{\Z})$ and $\bar{E}_2\in\Bor((\R^{p})^{\Z})$ uniquely determines events $E_{1}:=R_{1}^{-1}[\bar{E}_{1}]\in\E_{1}$ and $E_{2}:=R^{-1}_{2}[\bar{E}_{2}]\in\E_{2}$, respectively. Now, since \eqref{eq:rank_condition} holds, we have that the polynomial operator in the shift 
  \[
  R(\sigma):=\begin{bmatrix}R_1(\sigma)\\  R_2(\sigma)\end{bmatrix}
  \] 
  is a linear, continuous, and \emph{surjective} operator from $(\R^n)^\Z$ to $(\R^{m+p})^\Z$ \cite[\S4]{W:89}. Therefore, from 
  \begin{enumerate}[(i)]
  \item surjectivity of $R$, and 
  \item the fact that $R_{1}|_{(\R^n)^\Z/\ker_{\infty}R_{1}}$ and $R_{2}|_{(\R^n)^\Z/\ker_{\infty}R_{2}}$ are Borel isomorphisms,
  \end{enumerate}
   it follows that,  for any non-empty event $E_1\in\E_1$ and $E_2\in\E_2$, the intersection $E_1\cap E_2$ uniquely determines the set $E_1$ and $E_2$. (In particular, $E_1\cap E_2$ is non-empty, if $E_{1}$ and/or $E_{2}$ are so.)
   This in turn implies that $\E_1$ and $\E_2$ are complementary $\sigma$-algebras.

    \emph{``Only if''}: We prove the contrapositive, that is, if \eqref{eq:rank_condition} does not hold then $\E_1$ and $\E_2$ are not complementary. Assume that \eqref{eq:rank_condition} does not hold. Then the polynomial operator in the shift $R(\sigma):=[R_1(\sigma)^{\top}\  R_2(\sigma)^{\top}]^\top$ is not surjective, since the rows of $R(z)$ are linear dependent for every $z\in\C\setminus \{0\}$. Hence there exist two non-empty sets $E_1\in\E_1$ and $E_2\in\E_2$ whose intersection is the empty set. This in turn implies $E_1\cap E_2^{\mathrm{c}} = E_1=E_1\cap (\R^n)^\Z$. Therefore it follows that $\E_1$ and $\E_2$ are not complementary.\qed

\end{document}